\documentclass[12pt]{article}
\usepackage{psfig}
\usepackage{subfigure}
\begin{document}

\begin{center}
\LARGE{Globally and Locally Minimal Weight Spanning Tree Networks}
\vspace{0.2in}
\normalsize{}

A.R. Kansal$^1$ and S. Torquato$^{1,2}$
\end{center}

\noindent $^1$Department of Chemical Engineering, $^2$Department of
Chemistry and Princeton Materials Institute, Princeton University,
Princeton, NJ 08544

\begin{abstract}
The competition between local and global driving forces is significant 
in a wide variety of naturally occurring branched networks.
We have investigated the impact of a global minimization criterion versus 
a local one on the structure of spanning trees.
To do so, we consider two spanning tree structures - 
the generalized minimal spanning tree (GMST) defined by 
Dror et al. \cite{Dr00} and an analogous structure
based on the invasion percolation network, 
which we term the generalized invasive spanning tree or GIST.
In general, these two structures represent extremes of global and local
optimality, respectively.  Structural characteristics are compared 
between the GMST and GIST for a fixed lattice. In addition, we 
demonstrate a method for creating a series of structures
which enable one to span the range between these two extremes.  
Two structural characterizations, the occupied edge density (i.e., 
the fraction of edges in the graph that are included in the tree)
and the tortuosity of the arcs in the trees, are
shown to correlate well with the degree to which an intermediate
structure resembles the GMST or GIST. Both characterizations are 
straightforward to determine from an image and are potentially 
useful tools in the analysis of the formation of network structures.
\end{abstract}

\section{Introduction}

The purpose of the present research is to detail a new method by which 
information extracted from a single, fixed network structure can be 
utilized to understand the physical processes which guided the 
formation of that structure.  There are a variety of structures in 
nature and biology whose temporal development is difficult to observe.
Accordingly, the principle data from which to understand the 
physics that drove the formation of these structures is the final 
structure itself.

An excellent example of the use of a final network structure to
study the underlying physics is the work of Rodriguez-Iturbe
and Rinaldo on river basins \cite{Ro97}.  Detailed investigations of the 
structure of river basins combined with a variety of simulation
and theoretical analysis support the conclusion that minimal energy 
dissipation is the driving force (both global and local) in the 
structure of river basins \cite{Ri93,Su94,Ma96a}. Similarly,
natural complex branching patterns are observed in systems as 
diverse as retinal neurons \cite{Ka92}, dielectric breakdown 
\cite{Fr92} and human vasculature \cite{We97}.  
Another recent example is the network of invading cells in malignant 
brain tumors observed in vitro \cite{De01}.

All of these problems can be mapped to the language of spanning
trees.  For example, in the case of invading tumor cells, the
tumor cells form branched chains, i.e., tree structures.  The brain
offers these invading cells a variety of pathways they can invade
along (such as blood vessel and white fiber tracts) which may be
interpreted as the edges of an underlying graph, with the various 
resistances along these pathways playing the role of edge weights.
In many of these cases, the underlying physics behind the formation
of the observed patterns are only beginning to be understood. The
work presented here offers a useful tool in studying the driving
forces in the formation of these structures.

Here we consider the class of structures 
called \textit{spanning trees}.  Formally, spanning trees are 
defined on graphs and, in the most basic definition, are a loopless, 
connected set of edges that connect all of the nodes in the 
underlying graph (see Figure~\ref{fig:exmst}).
Many different spanning trees can be generated for any
given graph.  Therefore, it is possible to introduce
minimization criteria on the spanning-tree problem and select only those
trees which satisfy the criteria.  Thus spanning trees
represent an excellent test case for investigating the relation
between individual structures and the minimization criteria that 
govern their formation.

A broadly useful class of spanning trees (for examples see refs. 
\cite{Sc77,Ga83,Ad99}), 
is the minimal weight spanning tree (MST) \cite{Kr56,Pr57}.  
The MST is defined on an underlying graph whose edges each have some 
weight assigned to them.  The MST is then the spanning tree 
(a subset of the edges in the underlying graph) that 
minimizes the \emph{total} weight of the edges it includes.  The 
minimal weight spanning tree represents a structure whose formation 
is guided by a \emph{global optimization} principle.  It is also 
possible to define other types of criteria for spanning trees.  
For example, it is possible to define a spanning
tree such that only the lowest weight edges at each node are used
(a detailed discussion of such a class of structures follows), 
giving a system with purely local criteria.  Other types of
criteria can also be imposed, such as the degree constrained
minimum spanning tree \cite{Na80,Zh97}, but are not considered here.  

One of the structures we study in this paper is the generalized 
minimal spanning tree (GMST), proposed by Dror et al. \cite{Dr00}. 
The GMST is useful in considering problems in which there are
relevant length scales longer than a single edge.
For example, a biological system is characterized by the diameter
of a cell (mapped to a graph edge) as well as the length scale of 
diffusion in the system, which might be several cell diameters.
As the name suggests, the GMST is a generalization of the MST.
The GMST is defined on a graph in which the nodes have been
partitioned into groups. The spanning condition for the GMST is 
redefined (relative to the MST) such that instead of requiring that 
every node in the graph be included in the tree, the inclusion 
of at least one node from each group is required. The GMST 
structure is the tree that meets this definition of spanning and 
minimizes the total weight of the edges it includes.
When each group contains only one node, the GMST 
reduces to the MST.

The second class of trees considered in the present work 
is our generalization of the invasion percolation network 
\cite{Wi83} that we call the generalized invasive spanning tree 
(GIST).  The invasion percolation network begins with 
a connected cluster of edges (in the simplest case, this could be
just one edge).  This cluster then ``invades'' the remaining
edges by taking one edge from the boundary of the cluster and
including it in the cluster.  The edge that is included is the
single edge, of those on the cluster boundary, with the lowest 
weight. Additional edges are then included, one at a time, in the
same fashion until the cluster percolates (spans) across the system.
The generalization of the invasion percolation network to the 
GIST, in analogy to that of the GMST, partitions the nodes into
groups and modifies the percolation condition, such that one
node from each group must be spanned.

For graphs in which each group is a single node (i.e.
those graphs for which the GMST reduces to the MST), the GIST 
reduces to an acyclic invasion percolation network (i.e., 
an invasion percolation network without loops) \cite{Ba96}.  
It has been shown that the acyclic invasion percolation network
is identical to the MST \cite{Al95,Ba96}. Thus for graphs in 
which each group is a single node, the GMST and GIST are 
equivalent structures. Because of this equivalence, it 
is necessary to consider the GMST and GIST with groups of more 
than one node, rather than only the MST and invasion percolation, 
to understand the relation between local and global minimization criteria on 
tree structures.

The GMST and GIST structures were chosen because they generally 
offer extremes of global and local criteria.  Both classes of
trees have criteria which dictate 
the weight of edges chosen.  The GMST structures choose the
edges that minimize the total weight of the structure, even 
if that forces a higher weight edge to be chosen locally.
In contrast, the GIST structures include the lowest weight edge
locally, even if this results in a higher total weight for the
entire tree.  Except in the case of single-node groups noted above, 
each criterion results in a different final structure
(though by definition both yield spanning trees).  By comparing
these structures, the effect of each type of criteria can be 
identified.  Moreover, we provide a method to change a GIST structure 
incrementally into a more globally optimal GMST-like structure.
This allows various structural features to be observed as a function
of the degree to which either criterion is imposed.  These
intermediate structures can then serve as benchmarks for comparison
when a real image is analyzed.

This paper is organized as follows.  Section~\ref{sec:ma} is comprised of a 
description of the GMST and GIST structures.  It also contains 
brief summaries of the protocols used to form these trees, as well as
methods for transitioning from the GIST towards the GMST.  
Section~\ref{sec:bc} contains basic statistical descriptions of the 
structures generated for a given set of graph realizations.  
Section~\ref{sec:ad} introduces the statistical measurements of edge 
density and tortuosity and outlines how they may be used to study 
an experimentally observed image. Finally, Section~\ref{sec:co} has some 
concluding remarks.

\section{Models and algorithms}
\label{sec:ma}

Spanning trees were generated for graphs with 250,000 nodes arrayed
on a square lattice connected by 500,000 edges.  
The nodes were divided into groups of 64
nodes each, with each group containing the nodes within an eight by 
eight square.  The size of the groups was chosen such that the groups
were large enough to allow many possible subtrees within a group, but
still small enough to allow many groups within the graph.
Each edge was randomly assigned a weight uniformly 
distributed between 0 and 1.  Twenty graph realizations were generated 
using a different random seed for each (i.e., a different set of edge
weights).  The choice of an underlying graph that
conforms to the square lattice and groups that were spatially compact
was made to allow the resulting trees to be visualized readily.
The methods presented here, however, are general to any type
of graph.

We have employed one of the heuristics developed by Dror et al. 
for generation of the GMST structures \cite{Dr00}.  A brief summary 
of the algorithm that we have used is given below. The interested 
reader is referred to the original
paper for a more detailed discussion of several possible protocols 
for producing GMST's.  We have chosen to use heuristic H1 given 
by Dror et al.  For any given graph (generally containing many 
loops), this method begins with a small tree (initially comprised of 
a single edge) connecting two groups
via the most efficient path possible.  The tree is then ``grown'' by
adding the shortest path that enables the tree to span a new group 
(i.e., one not already spanned by the tree). 
This step is repeated until the tree spans all of the groups
in the system.

\noindent\textbf{GMST algorithm:}
\begin{enumerate}
\item Initialize the tree by choosing a group randomly and 
connecting it to the closest neighboring group.  Here closest 
means the group containing the node which can be added to the 
existing tree by the lowest weight path.  
\item Find the group closest to the tree not already included in 
the tree and connect it to the tree.
\item If all of the groups are now spanned, terminate the algorithm,
otherwise return to step~2.
\end{enumerate}

This algorithm was chosen because the specific details of our application
allow it to be implemented efficiently. For example, because the graph we 
use has a straightforward geometric interpretation, it is simple
to determine which group may be closest to the existing tree and what
nodes within the tree are likely to be close to these groups.  Thus,
rather than identifying the minimum path to each region at each
step, the distance to the majority of regions need not be calculated 
and further, for the regions that are checked, the distance from
only a limited number of the nodes included within the tree must be 
calculated.  Finally, as is discussed below, the size and regularity of 
the graph make it unnecessary to generate a distinct tree starting 
from each group and then find the tree with minimum total weight among 
these.  Instead, a very small sample of groups is used as possible 
starting points and the minimum of these is designated the GMST.

As a basis of comparison for the globally minimized structure that 
the GMST protocol generates, we have defined a class of structures 
which we term generalized invasive spanning trees, or GIST's. GIST 
structures meet the same spanning conditions as GMST structures, but
include local minimization in place of the GMST requirement of minimum
total weight.  The GIST structures are ``grown'' one edge at a 
time, analogously to the process of invasion percolation.

\noindent\textbf{GIST protocol:}
\begin{enumerate}
\item Begin with a tree comprised of the lowest weight edge
in the underlying graph.
\item Add to the tree the lowest weight edge remaining that adjoins
the existing tree.
\item If the newly added edge forms a loop with the edges already 
in the tree, delete it from the tree and from further possible 
additions.
\item If the tree now spans at least one node from each group, it
is the GIST.  Otherwise, return to step~2.
\end{enumerate}

The third step in the GIST protocol is included to 
ensure that a loop-less, tree structure is obtained, which is
analogous to the creation of ``trapped'' regions in the trapping
invasion percolation algorithm.  With this exception, 
the protocol outlined above is identical to a standard
invasion percolation protocol \cite{Wi83}.  
The only apparent difference is that
standard invasion percolation algorithms terminate once the
system is percolated (i.e., connected) in a few coordinate directions, 
rather than when it meets the spanning criterion given here. 
However, a change of language makes it clear that the spanning 
criterion is equivalent to a percolation condition of a rather
unusual structure in high dimensional space.  
Consider a graph consisting of $2K$ groups.  Each group of nodes 
can be mapped to one face of a hypercube in $K$ dimensions. 
A percolating cluster in this arrangement, which includes no internal
nodes, is one that connects all of the faces of the hypercube
or, in other words, one which connects at least one node from each
group.  Thus the group spanning requirement for a graph whose 
nodes are divided into $2K$ groups is equivalent to percolation 
of a hypercube whose nodes are confined to the faces of the cube in 
$K$ dimensional space.  Note that this equivalence
means that any of the highly-refined algorithms developed for
the generation of invasion percolation networks 
(e.g. Sheppard et al.~\cite{Sh99}) 
could be modified for use here affording tremendous 
computational speed-ups relative to the simple algorithm 
currently employed. 

Of importance for the purpose of comparison with GMST structures 
is that the criterion for adding an edge to the growing GIST 
structure is based only upon a local condition.  At every node in 
the system, the lowest weight edges will be included in the 
tree first, independent of which edge is most useful in meeting 
the ``terminating'' condition that requires all groups to be 
spanned. Thus the GIST represents the locally optimal extreme of 
all generalized spanning trees, while the GMST represents the 
globally optimal extreme. A careful comparison of the GIST and GMST 
protocols, however, reveals that under certain conditions the 
algorithms are identical.  Specifically, the GMST protocol 
reduces to the GIST one for the situation in which each group is 
of cardinality~1.  In other words, if each group is comprised of a 
single node, the closest new group is always connected to the 
existing tree by the lowest weight edge adjoining the tree, excluding
those edges that would form a loop. For such graphs the GMST and 
GIST reduce to the MST and the acyclic invasion percolation network, 
respectively.  The MST and acyclic invasion percolation structures 
have previously been shown to be identical \cite{Al95,Ba96}.  
As the groups increase in size, however, the GIST and GMST 
structures diverge from one another.

As reported below, the structures generated by the GIST protocol are
extremely dense by comparison to GMST structures and consequently
difficult to compare directly.  To address this issue we have adopted
a simple method termed ``pruning,'' which reduces the
density of the GIST.  In the pruning process, the edges comprising the
GIST are sorted in order of non-increasing weight.  Each ``redundant''
edge is then removed in turn, beginning with the highest weight edges.  
Here redundant is used in the same 
sense as in Dror et al., indicating an edge whose deletion
leaves a spanning tree \cite{Dr00}.  Indeed, this procedure is very 
similar to Heuristic H2 proposed by Dror et al. for the generation of 
a GMST from the MST of the same graph.  Because (unlike the heuristic
employed by Dror et al.) the initial tree in our use of the pruning 
algorithm is not an MST, the resulting tree is still quite far from 
the GMST. The result of repeatedly applying this algorithm to any 
initial tree is to reduce the density of the tree leading to a backbone 
structure.  This is similar to the concept of identifying the 
(elastic or flow-carrying) backbone of a percolation cluster.

Finally, a second transforming protocol was employed to study 
structures whose properties are intermediate between the GMST and 
the backbone of the GIST.  These intermediate structures serve as
useful comparison points for branched networks that are a
mixture of GMST and GIST backbone structures. Furthermore, investigating 
structures between the GIST backbone and the GMST allows us to assess
the value of different statistics over a range of structures, rather
than only at the extremal structures.  This method is an adaptation of 
the third heuristic proposed by Dror et al.  The procedure begins 
with any feasible spanning tree, which is then gradually converted 
to a more globally optimal structure.

\noindent\textbf{Conversion protocol:}
\begin{enumerate}
\item Begin with the backbone of the GIST.
\item Choose an arc (an unbranched chain of edges) within the tree.
\item Delete this arc and reconnect its end points with the path of
lowest weight.
\item Return to step~2, until a termination criterion is met.
\end{enumerate}

\noindent There are a few important observation that must be made 
regarding this protocol.  The most important is that it is not 
able to transform the tree structure to the same extent as the 
heuristic employed by Dror et al. \cite{Dr00}.
The main reason for this is the reconnection step (step~3) in this
protocol fixes the end points of the new arc, whereas Dror et al. 
allow the replacement arcs to begin and end at arbitrary nodes within 
the tree.  This is a compromise between the time complexity of the 
code and the efficiency of the resulting algorithm.  Without this
simplification, the time scaling of the code is too slow for use on the graph
sizes considered here.  In addition, the termination condition is also
chosen to reduce computation time.  The exhaustive search proposed
by Dror et al. is more effective overall, but requires too much
computational time to apply to the graphs used here.  Instead, we have 
used a condition in which arcs are chosen randomly and replaced until 
5000 successive replacements do not reduce the total weight of the 
spanning tree.

GMST structures took approximately 3 hours to generate on a single
node of an IBM SP2 computer.  GIST structures required roughly 4 minutes
each to generate.  Because of its comparative speed, the GIST routine 
was not optimized for execution time and could likely be accelerated 
considerably.  The run time 
of the pruning algorithm depends directly on the size of the tree 
in the input.  Runs on the initial GIST structures took approximately 
6 minutes each, while runs on structures close to the GIST backbone 
took less than one minute.  Finally, the conversion algorithm required 
approximately 10 hours per total conversion.

\section{Results and Standard Characterizations}
\label{sec:bc}

For each graph realization, GMST and GIST structures were generated.  
The GIST was then pruned repeatedly yielding a backbone structure, 
which was in turn reduced to a more globally minimal structure using 
the conversion protocol.  As noted previously, the GMST protocol
requires choosing a starting group.  While the best possible GMST
(within the limits of the heuristic method employed)
requires testing every group as a potential starting point, in
practice for our graphs this proved unnecessary.  To test the variation
in total weight caused by the choice of starting points,
the GMST protocol was run for a single graph 100 times with a 
different starting point each time.  In all
cases, the final structures were closely related to one another
(differing in less than 10\% of included edges).  Furthermore,
the total weights of the trees were narrowly distributed, with
a relative standard deviation of 0.2\%.  As such, almost any starting
group will yield a good approximation of the GMST.  For the data
presented here, we generated three potential GMST structures and
selected the minimum of those as the GMST.  Examples of the GIST 
and GMST are shown in Figure~\ref{fig:reals}.  The trees depicted 
are \textit{small samples of the entire tree structure}. Also shown 
in the figure are the trees resulting from the pruning and conversion
algorithms, which are labeled backbone and converted, respectively.
Note that the GIST has a much higher density than any of the other 
structures, making direct comparisons with it difficult.

Figure~\ref{fig:1comp} shows the total weights of several 
spanning trees for a single graph.  Two features from this 
figure merit special mention.  One feature is the gap in total weight 
between the result of the conversion protocol and the GMST. This 
difference highlights the limitations of the conversion protocol 
as formulated here.  In principle, a more complete conversion 
could produce trees closer to the GMST, for example by following the
protocol outlined by Dror et al. more faithfully.  However, as noted 
above, the computational cost of such an approach was prohibitive. 
While the protocol given above for the conversion of the backbone
structure calls for termination after a fixed number of consecutive
arc replacements leave the structure unchanged, the conversion process
may also be stopped after a fixed number of attempted replacements
or after a set total weight has been passed.  Either of these
choices will result in a structure that is only partially converted.
In other words, this will result in a structure that has some globally
dictated characteristics, but still has significant degrees of local
minimization incorporated.

The second notable feature is the large drop-off in total weight 
between the GIST and its backbone.  This drop indicates that a 
significant majority of the edges in the GIST structure (over 80\%) 
play no role in meeting the spanning requirement.  That many edges in 
the GIST can be removed is to be expected, in that edges are added to 
the GIST with no regard for the utility in creating a spanning tree. 
However, the degree to which edges can be removed is surprising.  
The pruning algorithm considers only the weight of an edge and
its role in maintaining a single, connected spanning tree.
This is a primarily local calculation and so it was originally 
expected that the backbone structure would be closely related to 
the GIST.  Instead, the majority of edges in the GIST are pruned 
away in creating the backbone.  

While the weights shown in Figure~\ref{fig:1comp} are for a
single graph, the variation in the weight of each tree between 
graphs is very small. This is not due to similarities in the actual 
trees, which display very few common edges.  For example, less that 
4\% of the edges in the GMST for one graph are present in the GMST 
for any other graph. Instead, it suggests that the total weight of
a GMST is insensitive to the specific details of the underlying graph.
The total weights of each type of tree averaged
over twenty different graph realizations are given in Table~\ref{tab:avg}.
Also listed in Table~\ref{tab:avg} are the average edge densities 
for each type of graph.  The edge density is calculated as 
the number of edges included in the spanning tree divided by the
number of edges in the complete graph. As expected, the edge density 
is strongly correlated with the total weight of each tree.
Because the minimization of the total tree weight is the objective
of the GMST protocol, the total tree weight can serve as an estimator
of the degree to which a tree resembles the GMST.

Figure~\ref{fig:wdist} shows the inclusion fraction distribution,
or more briefly the inclusion fraction, which is defined as the 
fraction of edges in the underlying graph of a given weight included 
in each tree as a function of edge weight. For the 
GIST, the inclusion fraction has a sigmoidal dependence on edge 
weight. Recalling the previous discussion of the equivalence between 
spanning and percolation, it is useful to compare the inclusion 
fraction of the GIST to that of invasion percolation networks.
The inclusion fraction for a sufficiently large (non-trapping) 
invasion percolation network is expected to take the form of a step 
function, with finite-size effects evidenced by deviations from 
the ideal function before and after the step \cite{Wi83,Wi84}.
The GIST distribution displays near perfect step behavior at
high edge weights, (with a vanishingly small number of high 
weight edges included) but a much more gradual drop-off at lower
edge weights relative to an ideal step function.  This behavior is
reminiscent of an invasion percolation network with trapping 
\cite{Wi83,Wi84}.  At large sizes, the inclusion fraction for 
this type of percolation network drops sharply beyond a critical
value of edge weight (like a step function). 
At low edge weights, however, the inclusion fraction drops off
gradually as the critical value is approached.  The area between
the inclusion fraction for the percolation network and the step
function approaches a non-zero constant for large systems.
This same type of low weight edge behavior is observed in the GIST
inclusion fraction.  A finite fraction of low weight edges are
not included in the GIST because adding them would create a loop.
In addition, there is a small area at the periphery of the graph 
(comprised of the most extreme groups) that is rarely visited by 
the tree.  This finite-size effect is the reason the inclusion 
fraction of the GIST does not reach its limiting value of one at 
the lowest edge weights.

The inclusion faction for the backbone structure matches 
that of the GIST exactly at high edge weights.  The few high weight 
edges included represent ``bottlenecks'' in the GIST, which are 
essential to the backbone. These bottlenecks should not be confused
with the most vital edge of a spanning tree, which is the edge
whose exclusion (and the subsequent reconnection of its end-points)
would cause the greatest increase in total tree
weight for a minimized tree \cite{Hs91}, though they are likely
to coincide to some degree.  Instead, they occur when, in the course
of creating the GIST, the invading tree reaches a point where all of 
the edges on the boundary of the current tree are high weight edges.
One of these high weight edges must be included in the GIST and cannot
be removed in the pruning process.  Based on our simulations, this 
situation arises several times in the typical construction of a GIST.
All of these instances are during the early stages of the tree formation,
however, and so a vanishingly small fraction of the high weight edges is 
included in the limit of an infinitely large graph. It is interesting 
to note that the inclusion fraction for the backbone structure 
has an extended plateau at low edge weights.  This suggests that low 
weight edges are equally likely to be included in the backbone structure
nearly independent of their weight (below a threshold).  
This can be contrasted with the curve for the GMST, in which 
lower weight edges are included more frequently than higher weight 
edges for all edge weights.  The inclusion fraction for
the converted structure is intermediate between the backbone and the
GMST.  Although the distributions shown in Figure~\ref{fig:wdist} differ
significantly from one another, the mean weight of the edges included
in each tree are very consistent (varying by a maximum of 0.02).

\section{Analysis and Discussion}
\label{sec:ad}

While the differences between the various types of trees in 
standard characterizations such as total weight or included edge 
fraction are clear, they are also of little utility in analyzing a
single given structure.  These measurements rely on the complete
knowledge of the graph, including all edge weights.  When
considering a physical problem such as the infiltration of tumor 
cells into a porous gel, this level of information is generally 
difficult, if not impossible, to obtain.  Instead a measurement 
relying only on the specific details of the observed structure and the 
most rudimentary information about the underlying graph is desirable.
Furthermore, in such examples the complete structure of the tree is
frequently difficult to image.  Accordingly, a statistical measurement
relying only on local information would be of particular value.

One measurement that meets these criteria is the occupied edge
density, which measures the fraction of edges in the underlying
graph that are included in the spanning tree, a measurement that
only requires an estimate of the total number of edges in the 
underlying graph (along with the network structure).  
As shown in Figure~\ref{fig:wdist}, the mean weight of
the edges included in each tree varies little between different
types of structures. Thus, the total weight of a tree
correlates strongly with the fraction of edges in the 
underlying graph occupied by the tree.  
Another useful measurement is the \emph{tortuosity}, $\tau$,
of the arcs in the tree.  Here tortuosity takes its common 
geometric meaning and is defined as an average of the ratio of the 
path length between two arbitrary nodes in the tree and the 
Euclidean distance between them.  For a relatively large tree, this
measurement can be made by averaging over the nodes included in the
tree (or even a portion of the tree, if the tree is large enough).
For small trees, however, an ensemble of trees would be 
necessary for an accurate measurement. This measurement can be made 
with no information about the underlying graph at all. 
For the trees considered here, the tortuosity is an increasing 
function of the path length.

\begin{equation}
\tau(\ell) = \left<\frac{\mbox{Path length between two arbitrary nodes}}%
{\mbox{Euclidean distance between nodes}}\right>
\end{equation}

\noindent In this equation, we have explicitly indicated that tortuosity
should be measured for a single path length $\ell$, hereafter
this notation will be suppressed and tortuosity will be indicated 
simply as $\tau$.
The angular brackets indicate an average over 
pairs of nodes (and over structures, if more than one is given).

Figure~\ref{fig:tort}(a) depicts average tortuosity 
curves for several types of spanning trees.  The variation in the
tortuosity curves between graph realizations is extremely small
(the relative standard deviation is less than 0.1\%) and as such 
is not indicated on the plot.  Note that the tortuosity curves
are ordered such that the GIST tree is the most tortuous, while
the GMST is the least tortuous, with the backbone structure and the
converted structure falling in between. Figure~\ref{fig:tort}(b)
shows the relation between total tree weight and tortuosity. In 
this figure, the average total weight of each type of tree is plotted
against the average tortuosity at a fixed path length. The length 
chosen here was twenty edge lengths, but similar plots
can be made for any path length up to the lengths considered.  
Tortuosity and total weight are positively correlated for all
of the trees considered.  The tortuosity and total weight of the 
spanning trees vary almost linearly between the backbone and GMST 
structures. The GIST structure, however, has a very high total
weight relative to the backbone structure that is not accompanied
by a correspondingly large increase in tortuosity.  This suggests
that in some respects the structure of the backbone resembles that 
of the GIST more closely than is indicated by the difference in
total weight.  Here it is simply noted that this resemblance comports 
with our initial expectation that the pruning algorithm would not 
affect the structure of the GIST to a very significant degree.
Further investigation is necessary to understand the relation
between a GIST and the resulting backbone tree more completely.

Assessing the degree to which a tree has been formed under a global
criterion requires more than just measuring the tortuosity (or 
occupied edge density).  In particular, the measurements presented
above are all specific to one type of graph - a square lattice with
random and uniformly distributed edge weights. Thus the scale set
by the GIST and GMST structures in the this work can only be used
to evaluate trees that develop on underlying graphs with 
similar average coordination and distributions of edge weights.  
If a tree develops on a graph that has a different distribution 
of edge weights or different coordination number, new standards for the 
tortuosity (for example) of the GIST and GMST structures
will need to be set.  A simple example will serve to illustrate
this point.  Table~\ref{tab:narr} compares the weight and tortuosity
of trees generated on several underlying graphs. The first data set is 
for the graph used in generating Figure~\ref{fig:1comp}. The second is 
for an identical graph except the edge weights are rescaled such that 
they span 0.45 to 0.55, while maintaining the same ordering.  Because
the GIST considers only relative weights (i.e., which edge has the 
lowest weight) this rescaling does not affect the structure of the 
GIST.  The GMST, however, is profoundly affected producing a tree that 
is significantly less tortuous at a fixed path length, though the total 
weight increases appreciably. Thus, a tree that developed on the 
second graph can have a very different balance of global and local 
influence than one developed on the first graph even though they both 
have the same tortuosity.  For example, while a tortuosity of 1.32 
would indicate a structure very close to the GMST of the first 
underlying graph, the same structure would have developed on the second 
underlying graph with a significant degree of local influence.

The need for standards (i.e. tortuosity measurements for the 
GIST and GMST) for the specific type of underlying graph 
presents a less serious obstacle than appears at first 
glance.  In particular, it is not necessary to reproduce an exact replica
of the real graph.  As discussed above, the variation of tortuosity 
between different realizations of the same type of graph is extremely
small. Thus, the only knowledge required to measure and evaluate 
the tortuosity of a tree is a statistical understanding of the 
underlying graph.  In addition, computational
experiments indicate that the exact form of the edge weight 
distribution also does not impact the tortuosity of the GMST 
significantly.  To make this assessment, we have generated a GMST
structure on a graph with edge weights drawn from a Gaussian 
distribution with mean 0.5.  The Gaussian distribution
was scaled such that the standard deviation matched that of a
uniform distribution between 0 and 1, with values below 0 or above 1 
set to the appropriate extreme.  The statistical properties of the
GIST and GMST on this graph are listed in Table~\ref{tab:narr} as 
Graph~3. The results are very similar to those for a graph with uniformly
distributed edge weights.  While it is necessary to test several
other types of edge weight distributions before the claim can be
confirmed, these results indicate that tortuosity is not sensitive
to the exact form of the edge weight distribution. Recalling
that the structure of the GIST is not determined by the distribution
of edge weights, this means that setting the tortuosity standards
for a general class of underlying graphs can be accomplished with 
very limited information about the specific graph.

In practical terms, given a single tree structure (or an ensemble
of small structures), a simple procedure can be followed to estimate
its relation to the GMST and GIST.

\noindent\textbf{Evaluation protocol}
\begin{enumerate}
\item Measure the tortuosity, $\tau$, for the given tree at a fixed
path length.
\item Generate the GIST and GMST for the underlying graph thought to 
exist in the problem being considered and measure their tortuosities.
\begin{enumerate}
\item Estimate the coordination number of the expected underlying 
graph. Generate a model underlying graph, with the same type of 
coordination.  For example, expecting a coordination number of 6, one
could use a triangular lattice.  Estimate the length scale of any
long-range effects in the system and group the nodes in the
underlying graph in accordance with this length scale.
\item Using \textit{any} distribution of edge weights, generate the
GIST and measure its tortuosity, $\tau_{GIST}$.
\item Estimate the relative dispersion of the edge weights in the 
graph. This requires an approximation of the ratio $\sigma/\mu$, 
where $\sigma$ is the standard deviation and $\mu$ is the mean
of the edge weights.  
\item For the underlying graph chosen in step 2(a), generate a graph
realization whose edge weights are drawn from a distribution with the
relative dispersion estimated in step 2(c).  Generate the GMST for this
graph realization and measure its tortuosity, $\tau_{GMST}$.
\end{enumerate}
\item Evaluate the ratio 
\begin{equation}
\overline{\tau} = \frac{\tau-\tau_{GMST}}{\tau_{GIST}-\tau_{GMST}}
\end{equation}
\noindent which represents the degree to which the given tree 
resembles the GIST. For example, a value of $0.1$ would indicate a
structure that is primarily globally optimal (GMST-like), while
$0.8$ would indicate a structure that is dominated by local effects
(GIST-like).
\end{enumerate}

As an example, consider the problem of cells moving through a porous 
medium under the influence of a nutrient gradient.  The pores in the 
medium play the role of edges in the graph and their intersections
are the nodes.  Assume that we can measure the coordination number of
the pore structure, perhaps by the same imaging technique that has
produced the tree we are analyzing.  For a medium with a coordination
number of three, we might choose a hexagonal lattice as our graph.
The long-range effect we are investigating is the influence of the 
nutrient gradient, so groups in the graph should be the size of the 
nutrient diffusion length scale.  So our underlying graph would now
be a hexagonal lattice, tiled into groups each of which has the same
length scale as the nutrient diffusion length in the real system.
Using this graph, we can use any edge weight distribution we choose
and generate the GIST.  To generate the GMST, however, it is necessary
to make one more estimate - we must decide what type of resistances
are present.  For this problem, the resistance might be caused by
the cells squeezing to fit into the pores. Thus the weight of an edge
would be the inverse of the pore diameter.  After estimating the 
range of pore diameters in the porous medium, we can construct an
edge weight distribution that matches this estimate.  Using this 
distribution, we can complete our graph realization and measure the
tortuosity of the GMST.  Note that the only information required in
this process was very general and should be relatively easily 
accessible experimentally.

\section{Conclusions}
\label{sec:co}

The results described above show that altering the criterion for
including edges in a spanning tree from a global one (the GMST) to
a local criterion (the GIST) has a measurable impact on the 
statistical characterization of the resulting trees.  In particular,
the total weight, occupied edge density, inclusion fraction distribution, 
and the tortuosity varied systematically with different types of spanning
trees.  Of these, however, the majority require extensive information
about the graph under consideration, including individual edge weights.
In contrast, however, measurement of the tortuosity of a tree only 
requires information about the structure of the tree.  
Measurement of the occupied edge density requires minimal information
about the underlying graph, in addition to the tree structure,
but does not require the detailed information necessary for measurements
like the total weight or the inclusion fraction distribution.  
As noted above, both the edge density and the tortuosity
measurements for a single network must be made in the context of the
\textit{type} of underlying graph present.  We emphasize that while
this context requires some basic knowledge of the characteristics of
the underlying graph, this is a much more accessible level of information
than the complete information required for measurements such as the total
weight of the tree.

Using tortuosity to characterize a network has one additional 
advantage. The tortuosity is measured along individual arcs for 
relatively small path lengths.  Thus the tortuosity of a structure can 
be measured accurately even if only a small portion of the structure
is observed.  This is of particular value in assessing systems that
are challenging to image completely.  These features make the 
tortuosity and occupied edge density promising tools for the 
investigation of naturally occurring tree structures whose temporal 
formation cannot be observed directly.

One type of variation that is not addressed in the discussion above
is a change in the coordination number of the graph.  It has 
been shown that changing the coordination of the underlying graph 
has a pronounced
impact on the percolation threshold in the invasion percolation system
\cite{Wi83}.  As such, it is expected that similarly
pronounced impacts would be observed on the structure of any spanning
tree.  Characterizing the effect of different coordination numbers 
on the structure of the GMST, however, follows the
exact methodology outlined in this paper.  In sum, employing the
protocols discussed previously on a graph of interest will yield
a set of benchmark tortuosity or occupied edge density measurements 
that may then be used to assess whether global or local weight 
criteria played an important role in the development of 
any given tree structure.

\section*{Acknowledgements}
This work has been supported in part by grants CA84509 and 
CA69246 from the National Institutes of Health.
The work was also supported by the Engineering Research 
Program of the Office of Basic Energy Sciences at
the Department of Energy (Grant DE- FG02-92ER14275).
The authors would like to thank Dr. T. S. Deisboeck for valuable 
discussions.

\bibliographystyle{unsrt}

\newpage

\section*{Figure Captions}

\begin{figure}[htb]
\centerline{\subfigure[]{\psfig{file=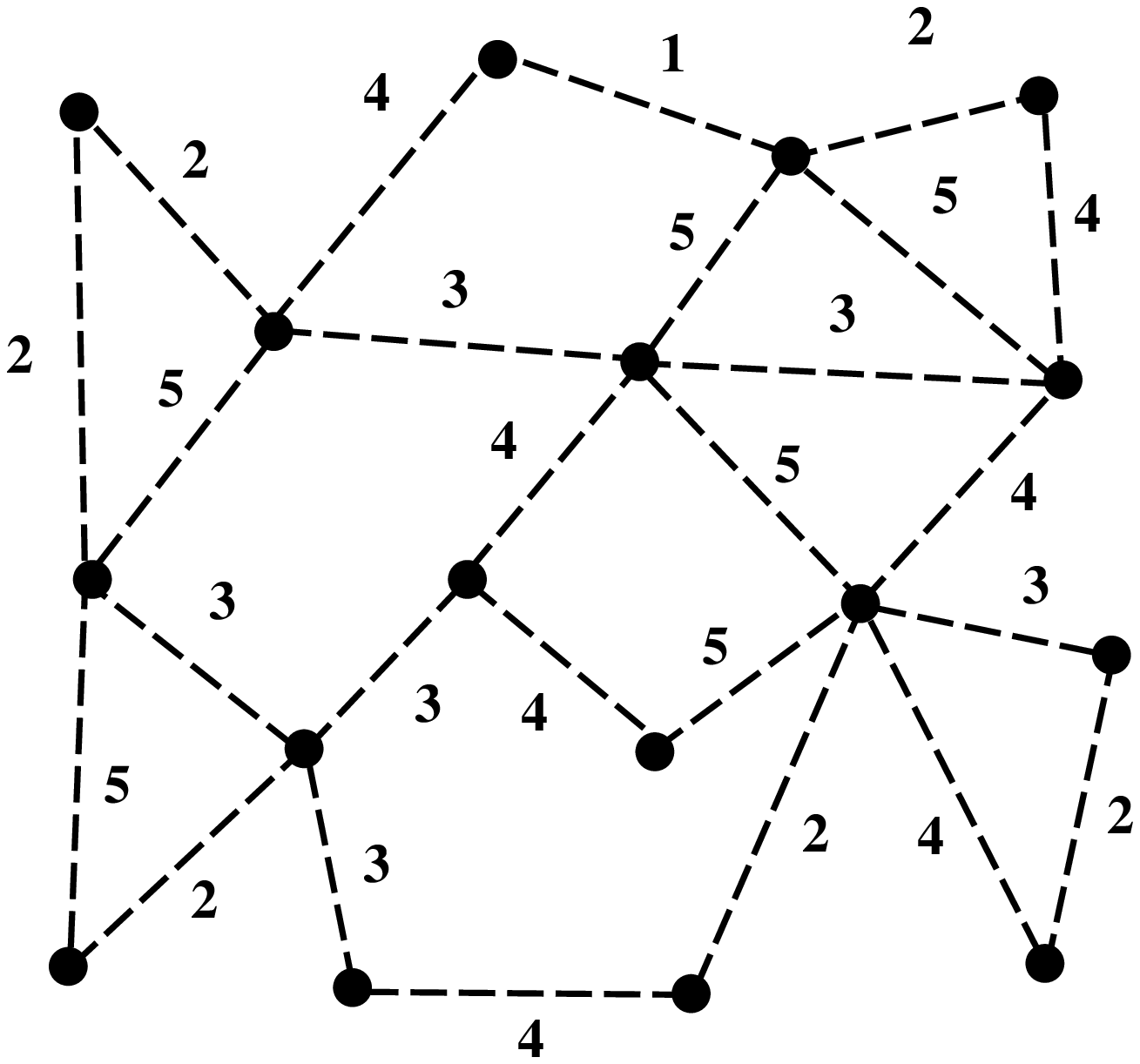,height=2in}}} 
\centerline{\subfigure[]{\psfig{file=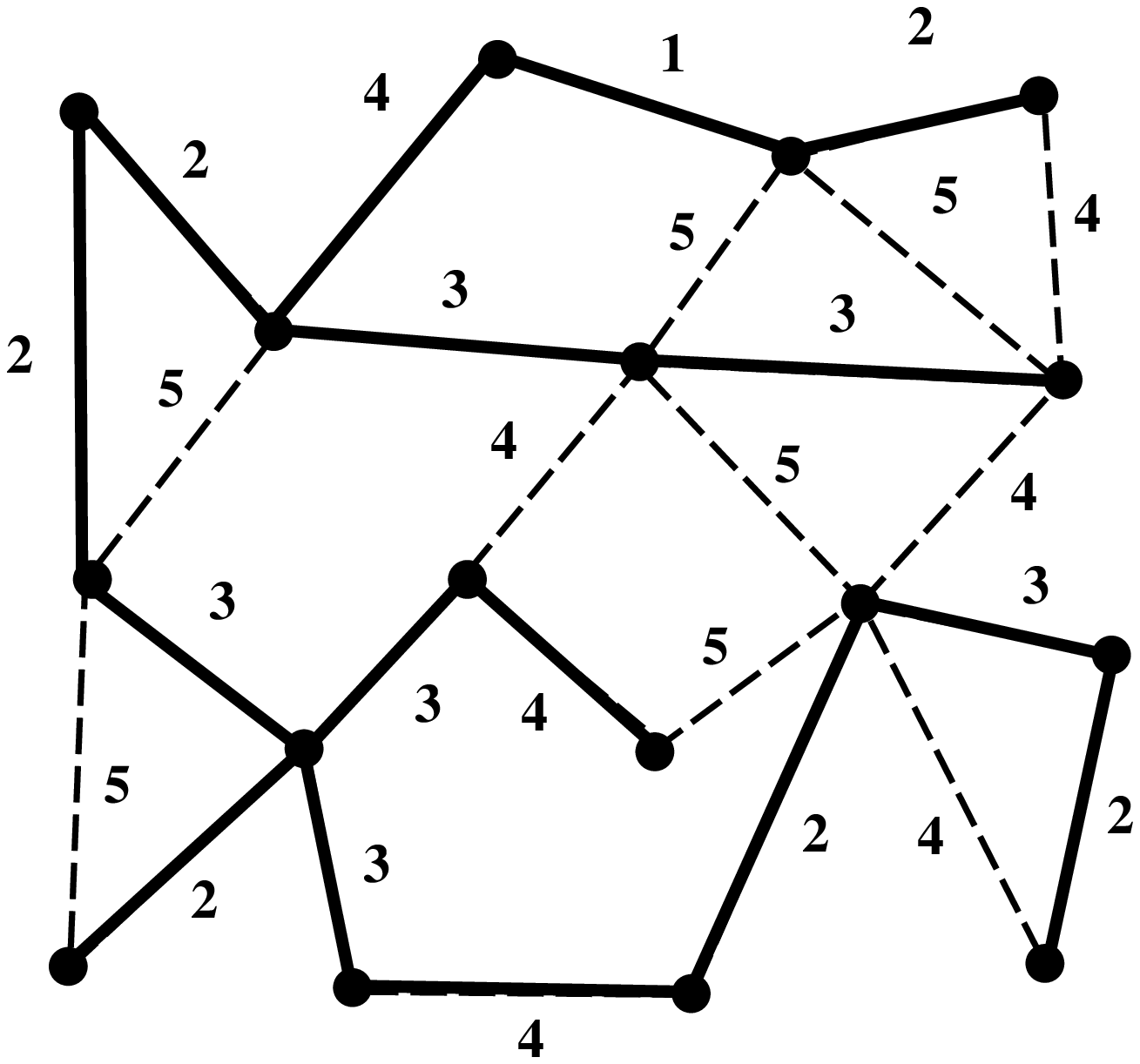,height=2in}}}
\centerline{\subfigure[]{\psfig{file=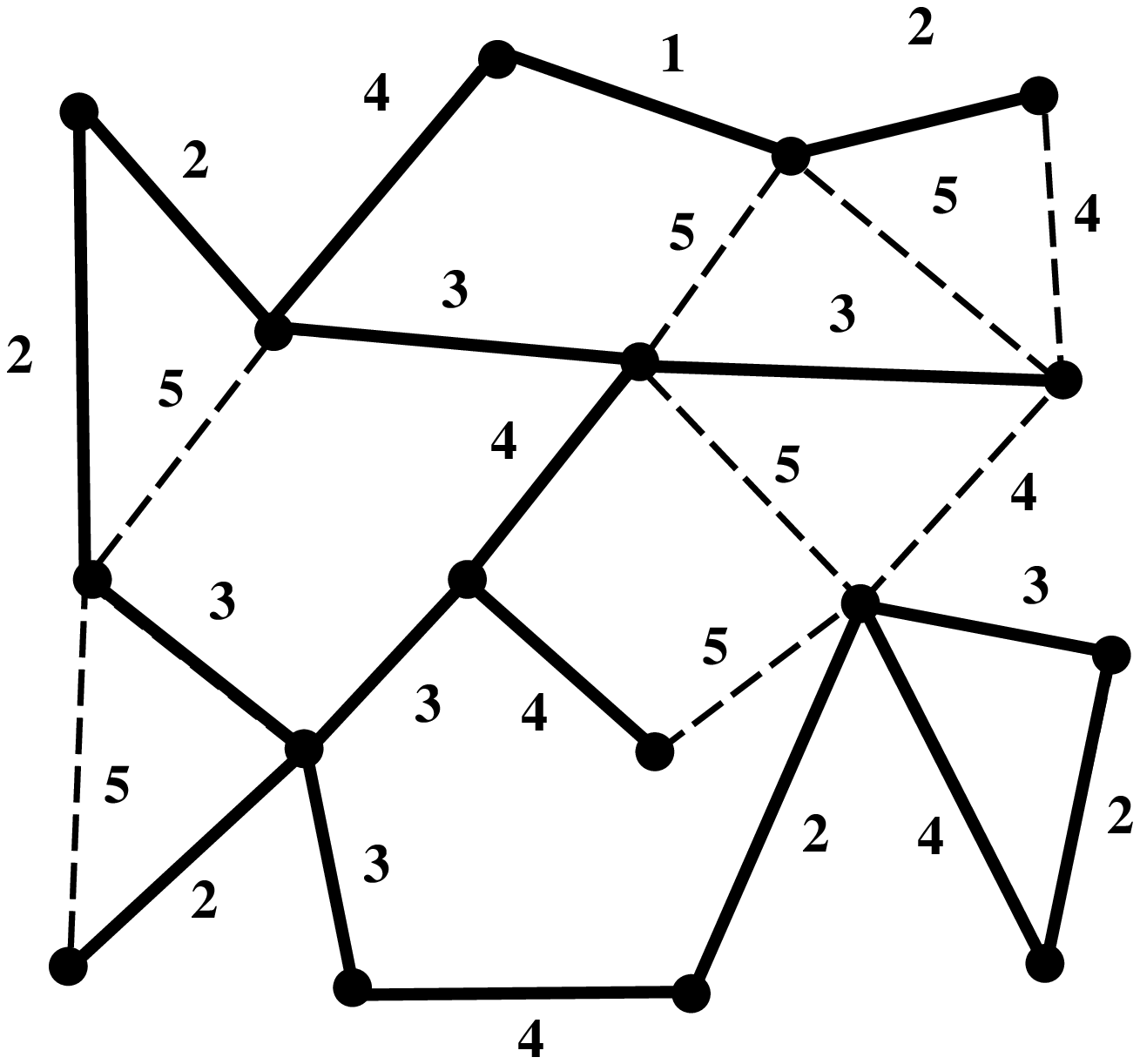,height=2in}}}
\caption{}
\label{fig:exmst}
\end{figure}

Figure~\ref{fig:exmst}:  Example of a weighted graph and the 
resulting minimal
spanning tree.  (a)~shows all of the edges and nodes in a graph,
with the weight of each edge indicated next to the edge.
Graph edges are depicted by broken lines.
(b)~shows the minimal spanning tree for this graph, which is the
set of edges that connects every node in the graph in the tree with
the lowest total weight. Edges included in the tree are shown as 
solid lines, while edges not included remain broken lines. The total
weight of the tree in (b) is 40, and the occupied edge density 
(number of edges included in the tree divided by total number of edges
in the graph) is $15/25 = 0.6$. (c)~shows the invasion percolation network
for the same graph.  Note that the invasion percolation network may have
loops and in this case there are two closed loops.  If loop formation is
prevented (resulting in the highest weight edge in any loop remaining
unoccupied) the result is the acyclic invasion percolation network.
As can be readily seen by comparing figures (b) and (c) the acyclic 
invasion percolation network is identical to the MST.

\begin{figure}
\centerline{
\subfigure[]{\psfig{file=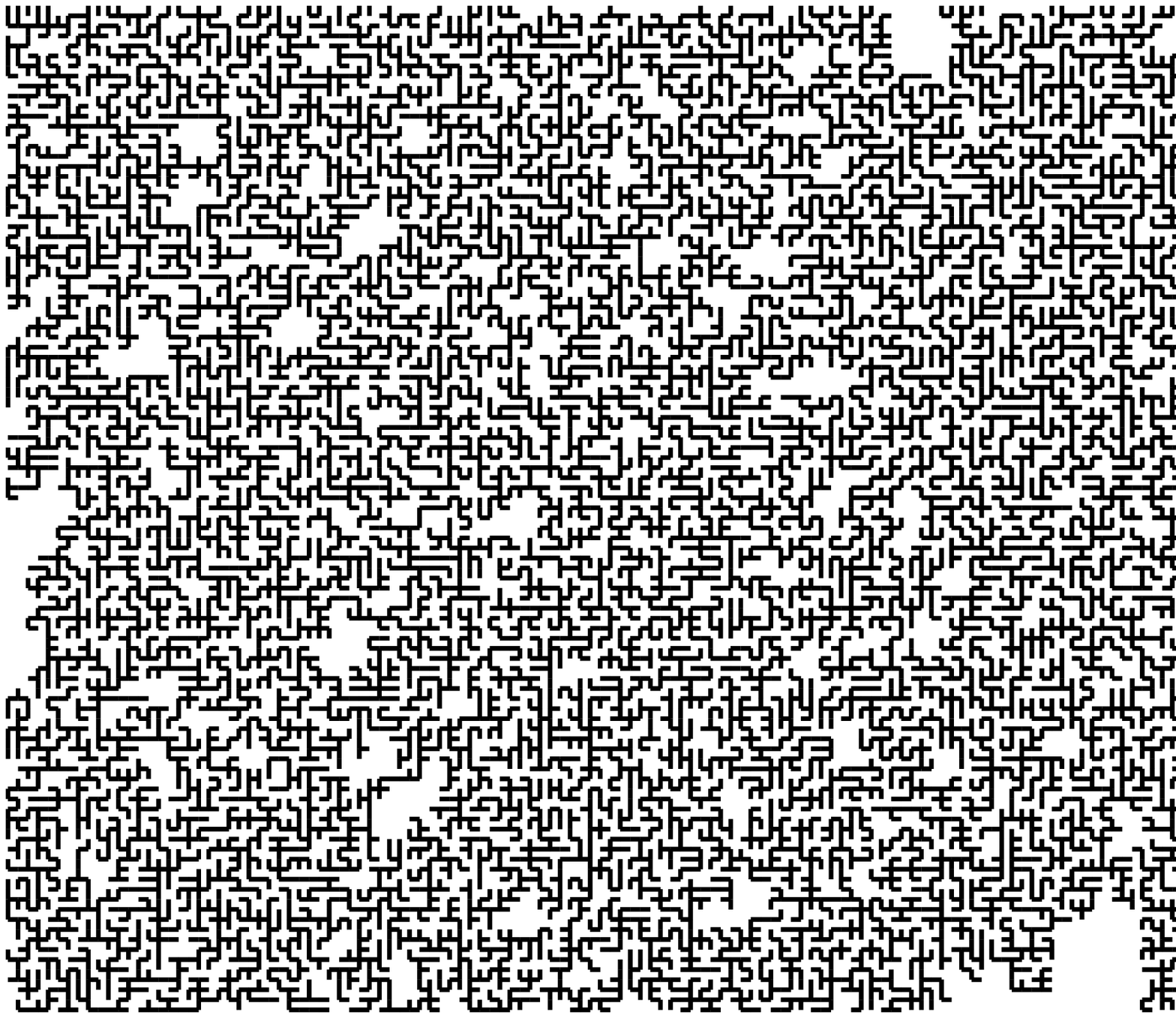,width=0.4\textwidth,clip=}}
\hspace{0.5in}
\subfigure[]{\psfig{file=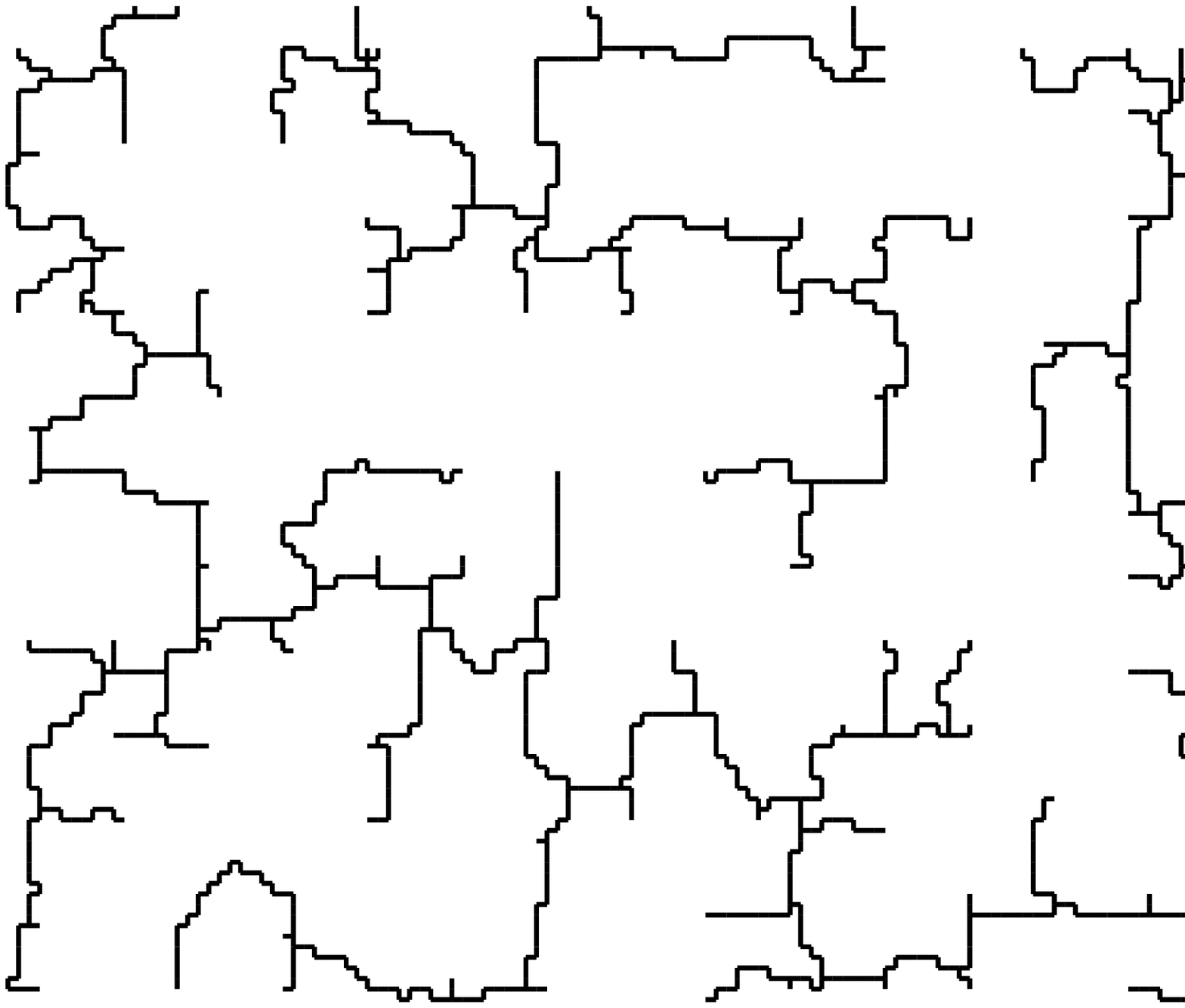,width=0.4\textwidth,clip=}}}
\centerline{
\subfigure[]{\psfig{file=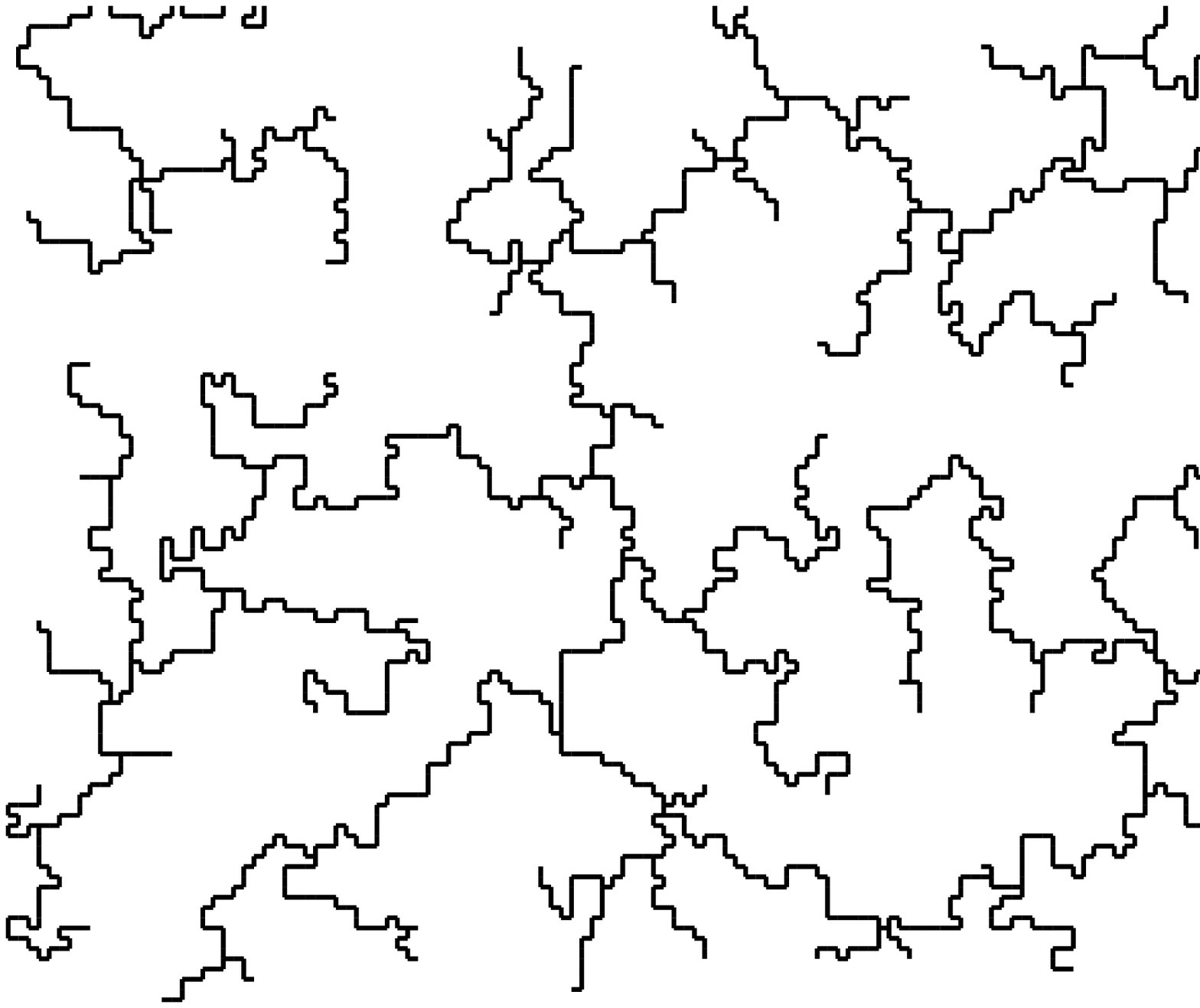,width=0.4\textwidth,clip=}}
\hspace{0.5in}
\subfigure[]{\psfig{file=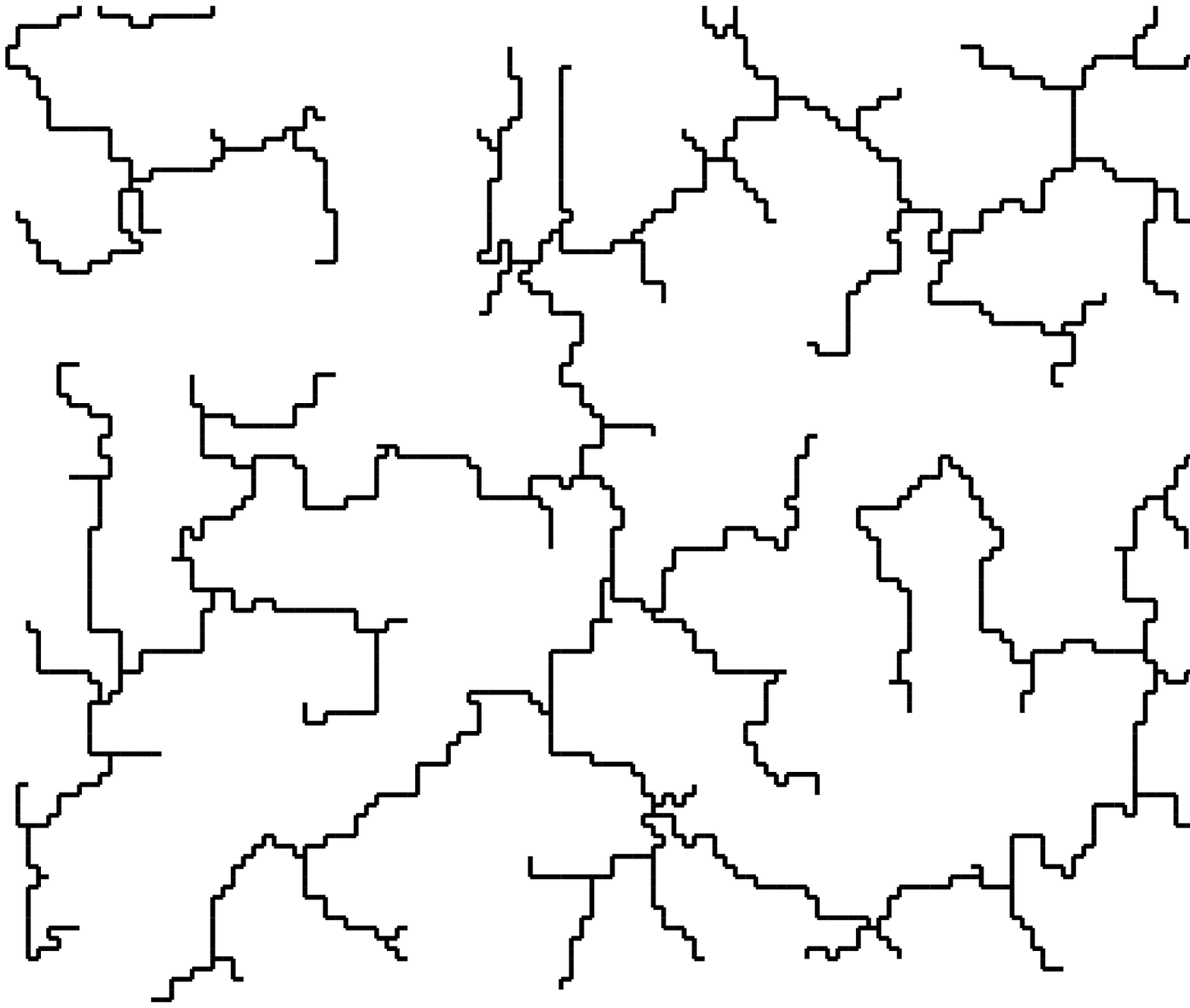,width=0.4\textwidth,clip=}}}
\caption{}
\label{fig:reals}
\end{figure}

Figure~\ref{fig:reals}:  Examples of the (a)~GIST, (b)~GMST, (c)~backbone, 
and (d)~converted structures for a single graph.  The backbone structure 
is the result of repeatedly applying the pruning algorithm to the GIST.
The converted structure is the result of applying the conversion
algorithm to the backbone structure.  In each case a small, 
representative section is shown (not the entire tree). The underlying
graph is not shown in these images.

\begin{figure}[htb]
\centerline{\psfig{file=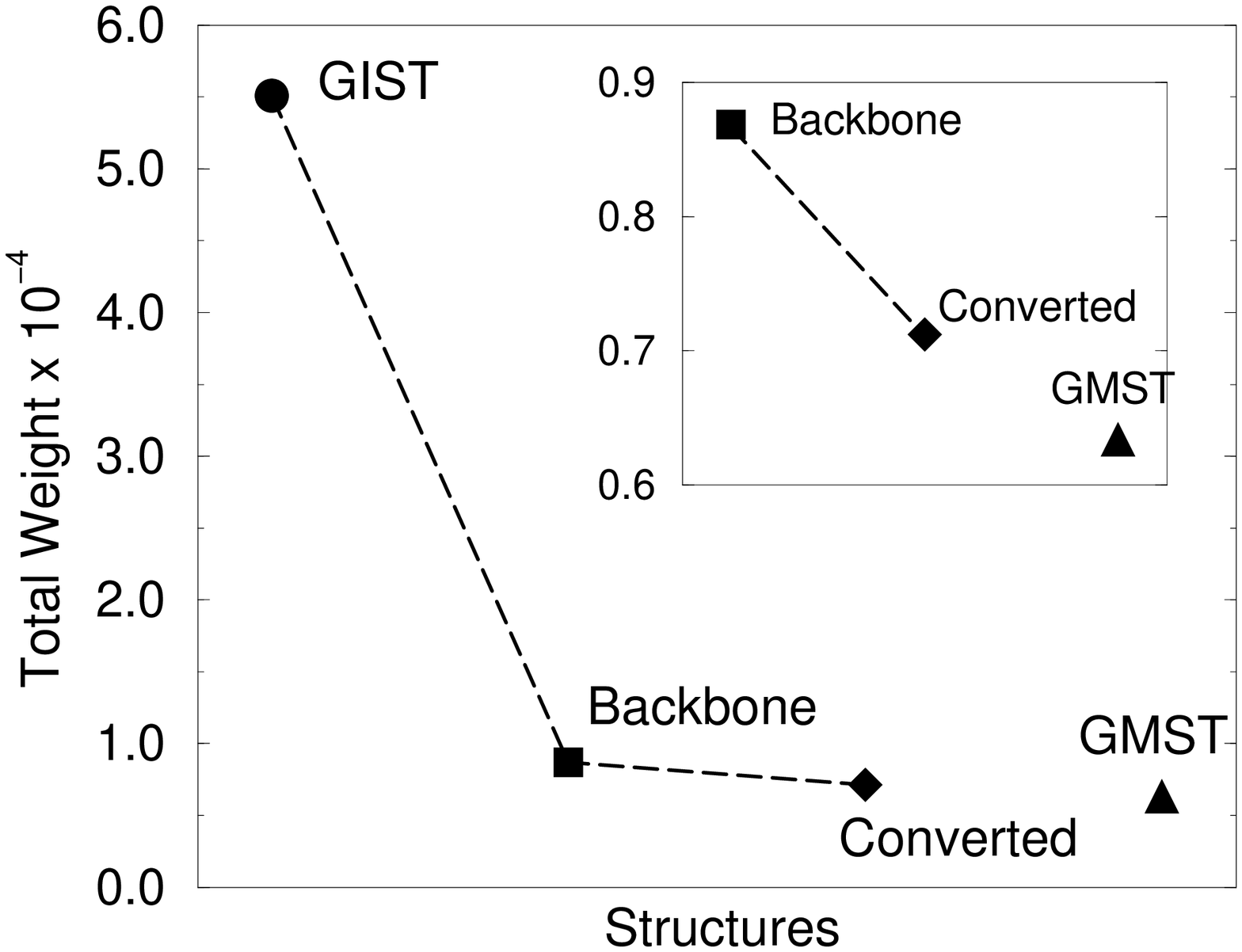,height=2.5in}}
\caption{}
\label{fig:1comp}
\end{figure}

Figure~\ref{fig:1comp}:  The total weight of several types of trees 
for a single graph realization. Inset is a magnification of the 
three low weight trees to highlight the relative difference.
The GIST, backbone, and converted structures fall along
a continuum of structures as indicated by the connecting line.  
The backbone structure is generated by repeatedly pruning the GIST.  
The converted structure is the result of applying the conversion 
protocol to the backbone structure. Spanning trees between the 
converted tree and the GMST have yet to be produced.

\begin{figure}[htb]
\centerline{\psfig{file=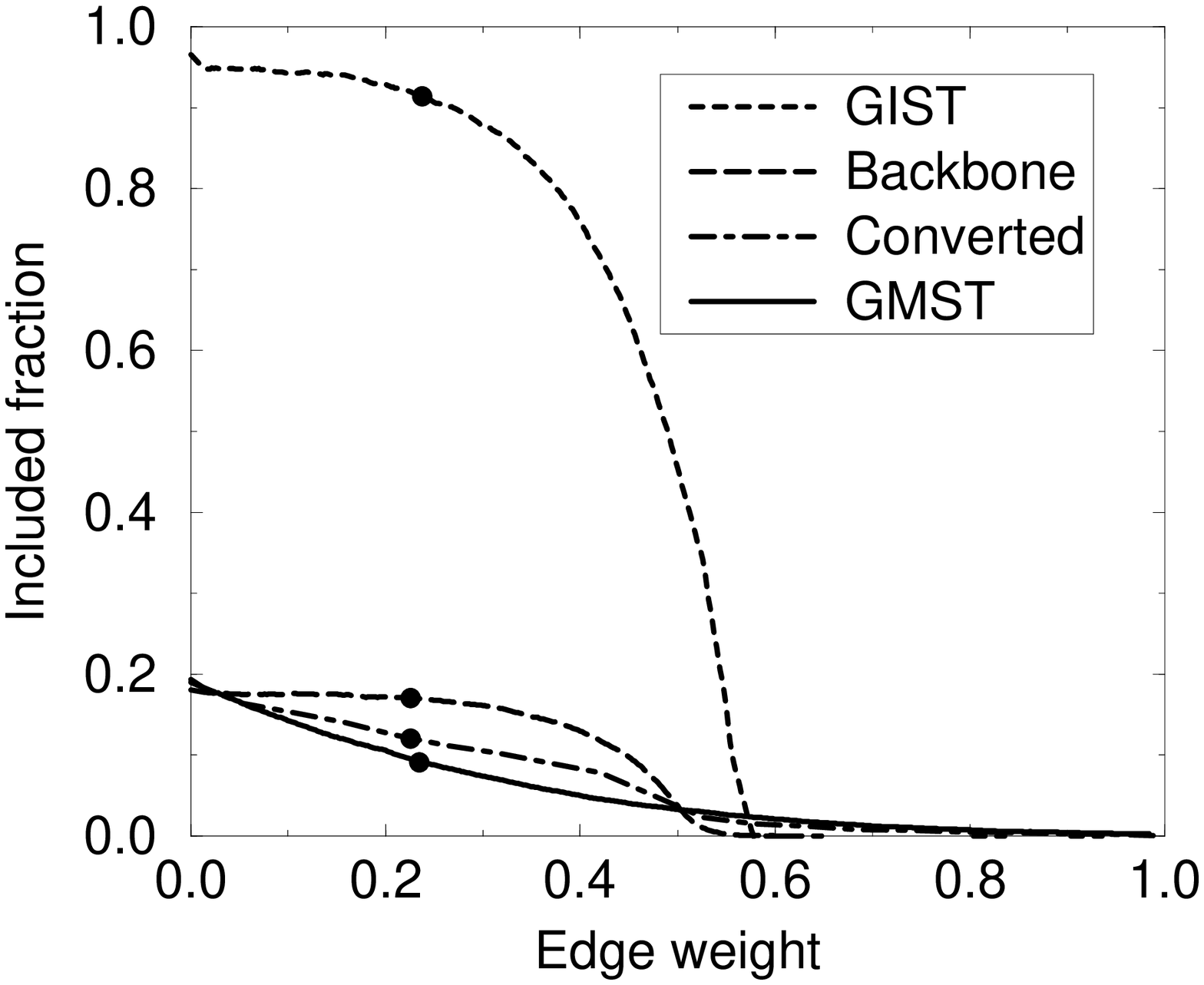,height=2.5in}}
\caption{}
\label{fig:wdist}
\end{figure}

Figure~\ref{fig:wdist}:  Plot of the fraction of edges included in each tree 
as a function of edge weight.  The mean weight of the edges included
in each tree is indicated by the circle on each distribution.

\begin{figure}[htb]
\centerline{\subfigure[]{\psfig{file=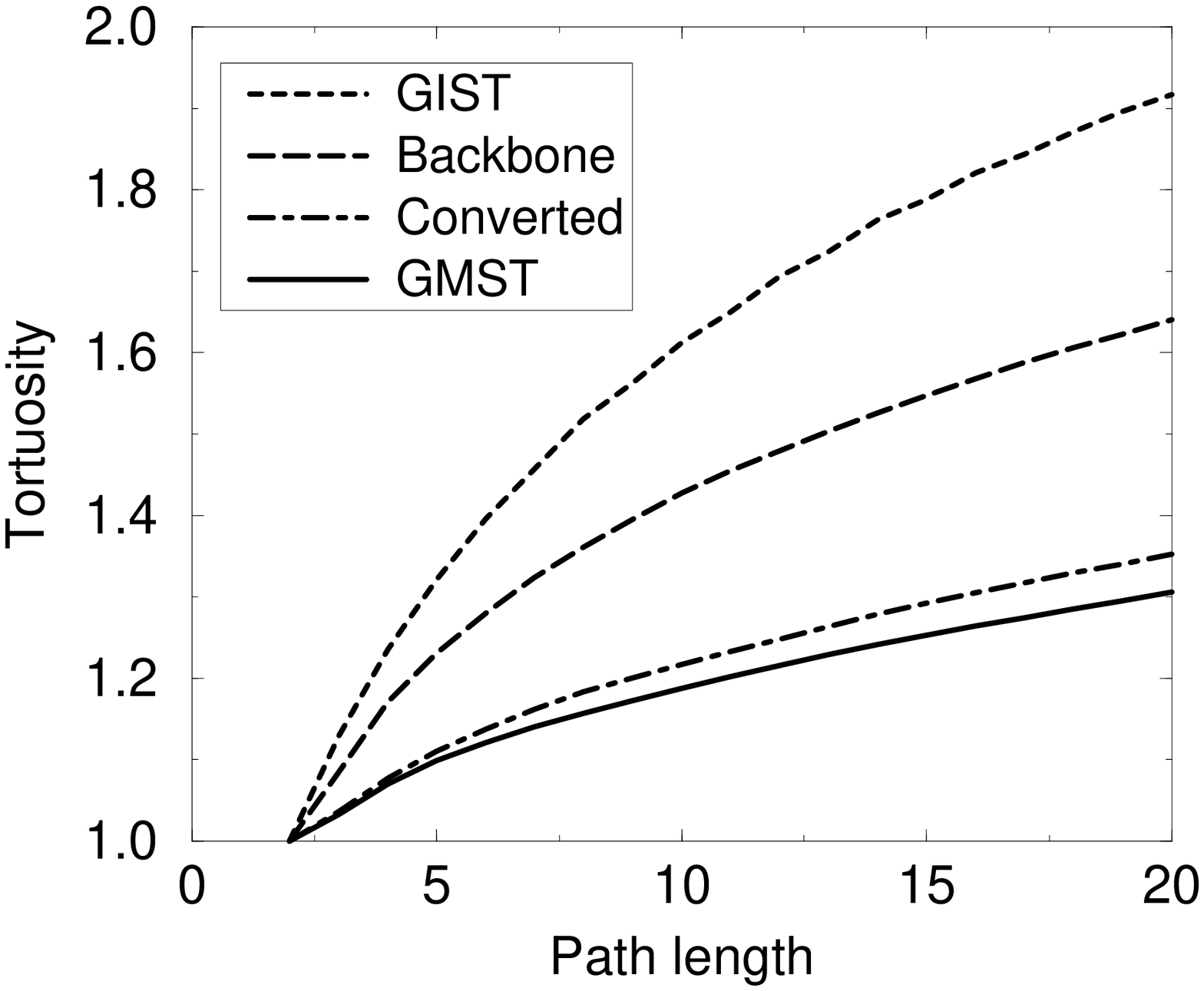,height=2.5in}}
\hspace{0.25in}\subfigure[]{\psfig{file=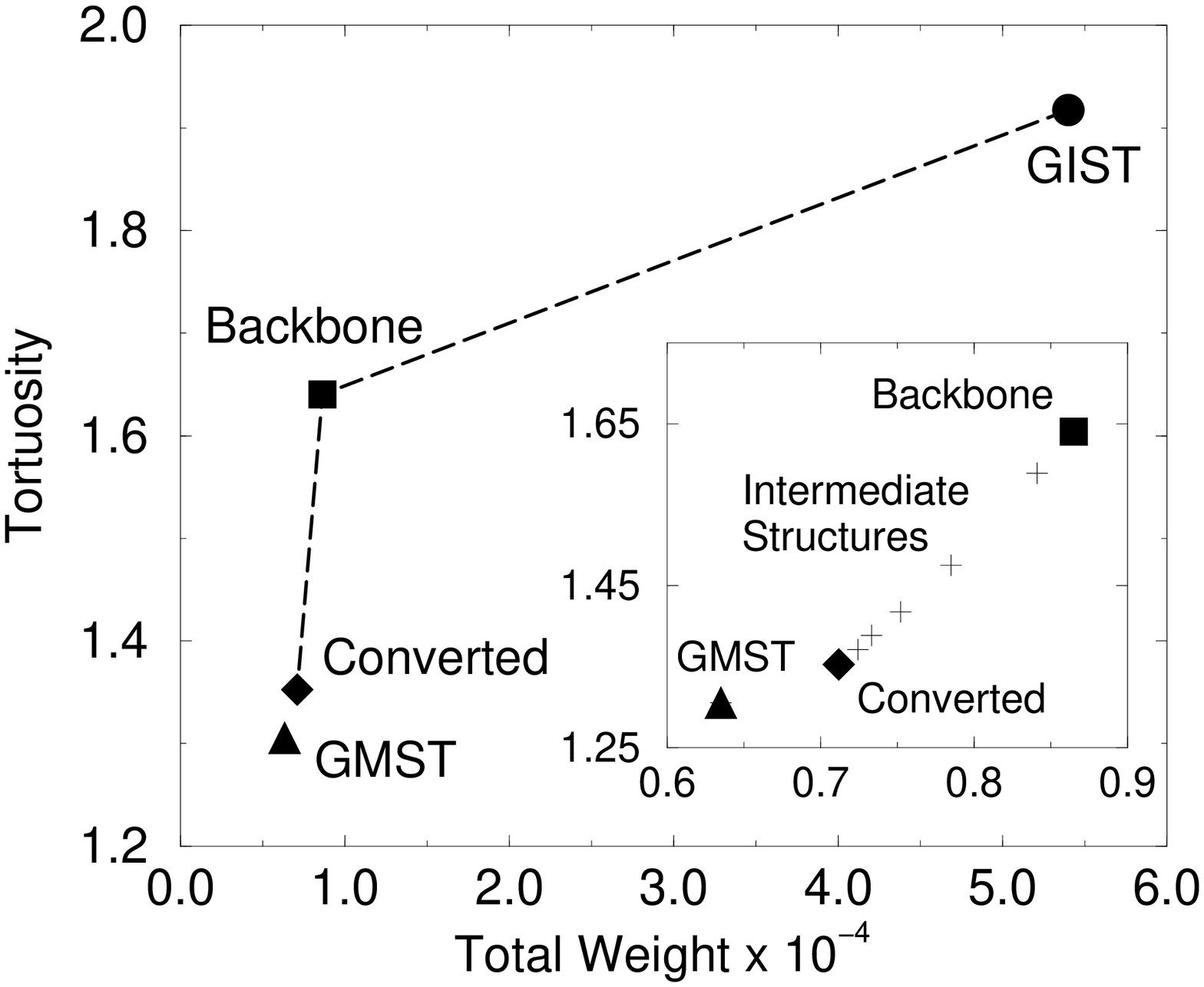,height=2.5in}}}
\caption{}
\label{fig:tort}
\end{figure}

Figure~\ref{fig:tort}:  
(a) Tortuosity versus path length for several types of trees
averaged over twenty graph realizations.  The variation between
graphs is sufficiently small that the curve for each graph is 
indistinguishable from the averaged curve.
Note that the ordering of the curves corresponds to the total tree 
weights (i.e., high tortuosity correlates with high total weight).  
This correlation is explicitly displayed in (b). Inset is a 
magnification of the low total weight points, emphasizing their 
relative differences.  Also included in the inset are several
intermediate structures that are generated during the conversion
algorithm.

\newpage

\begin{table}[htb]
\caption{Total weight and edge density of different 
types of spanning trees averaged over twenty graphs. As 
indicated in Figure~\ref{fig:1comp}, the backbone structures
result from pruning the GIST, and in turn are used to produce 
the converted trees.}
\label{tab:avg}

\begin{tabular}{l|rl|rl} \hline
Structure & \multicolumn{2}{c|}{Weight} & \multicolumn{2}{c}{Edge density} \\ 
\hline
GIST & 54000&$\pm$ 1400 & 0.4539&$\pm$ 2.1$\times 10^{-2}$\\ \hline
Backbone & 8650&$\pm$ 67 & 0.0766&$\pm$ 4.9$\times 10^{-4}$\\ \hline
Converted & 7120&$\pm$ 82 & 0.0630&$\pm$ 6.2$\times 10^{-4}$ \\ \hline
GMST & 6350&$\pm$ 35 & 0.0541&$\pm$ 2.3$\times 10^{-4}$\\ \hline
\end{tabular}
\end{table}

\newpage

\begin{table}[htb]
\caption{Comparison of the GIST and GMST on three different graphs.  
All three graphs are square lattice, with different edge weight 
distributions. The tortuosity is measured at a fixed path length of 20.
The second graph is identical to the first, except its edge weights 
have been rescaled to lie between 0.45 and 0.55 (versus 0 and 1.0 in 
the first graph).  The edge weights in the third graph have been 
converted to follow a Gaussian distribution with the same mean and 
standard deviation as the first graph.
The rescalings do not affect the structure of the GIST at all,
though the total weight can change.  
In contrast, the GMST structure is dependent of the edge weight 
distribution used.  However, comparison of the first and third graphs
reveals that there may not be a strong dependence on the exact form
of the distribution.}
\label{tab:narr}

\begin{tabular}{l|l|r|r|} \hline
\multicolumn{2}{c|}{~} & Weight & Tortuosity \\ \hline
Graph 1 & GIST & 55080 & 1.9175 \\ \cline{2-4}
& GMST & 6341 & 1.3055 \\ \hline
Graph 2 & GIST & 107637 & 1.9175 \\ \cline{2-4}
& GMST & 10239 & 1.1365 \\ \hline
Graph 3 & GIST & 58361 & 1.9175 \\ \cline{2-4}
& GMST & 6350 & 1.3101 \\ \hline
\end{tabular}
\end{table}

\newpage

\end{document}